\documentstyle[12pt,twoside]{article}
\input{epsf}
\begin{document}
\setlength{\baselineskip}{12pt}
\hoffset 0.65cm
\voffset 0.3cm
\bibliographystyle{normal}

\vspace{48pt}
\noindent
{\bf Drifting Abnormal Rolls in Electroconvection of Hybrid\\ Aligned
Nematic}

\vspace{48pt}
\noindent
V.~A. DELEV$^a$ and A.~P. KREKHOV$^{a,b}$

\noindent
$^a$Institute of Molecule and Crystal Physics, Russian Academy of
Sciences, 450025 Ufa, Russia;
$^b$Physikalisches Institut, Universit\"at Bayreuth,
D-95440 Bayreuth, Germany

\vspace{36pt}
\noindent
We report experimental and theoretical results on the conductive regime of 
electroconvection in hybrid aligned nematics. The drifting 
oblique/normal rolls below/above the Lifshitz frequency are observed at the 
onset of electroconvection under a.c. voltage. The experimental 
data on the threshold voltage, wavelength, oblique angle and drift
period of the rolls as function of a.c. frequency are in good 
agreement with the results of linear stability analysis. The transition from
drifting oblique rolls to abnormal rolls is found below
the Lifshitz frequency with increasing applied voltage.

\vspace{24pt}

\vspace{36pt}
\baselineskip=1.5 \baselineskip

\noindent
{\bf INTRODUCTION}
\vspace{12pt}

\noindent
Electroconvection (EC) in nematic liquid crystals (NLCs) has been
studied extensively in experiment and theory for about 30 years.
Depending on the alignment of the director $\bf\hat{n}$ that gives 
the preferred 
direction of molecular 
orientation, the EC system demonstrates 
a rich variety of pattern-forming instabilities 
(see, e.g., \cite{KP_Book96}).
In particular, in planarly aligned nematics the rolls at onset are oriented 
either normally to the initial director orientation (normal rolls) 
above the Lifshitz frequency $f_L$ (crossover from oblique to normal rolls at
threshold) \cite{KH_MCLC77, RJL_PRL86} or obliquely to it (oblique rolls) 
below $f_L$ \cite{RJL_PRL86, BZK_JP88}.
With increasing applied voltage one observes 
various secondary instabilities, such as zig-zag and 
skewed-varicose patterns (see \cite{KP_Book96} for a review).
Recently, a new secondary instability leading to abnormal rolls (ARs) that 
are characterized by a homogeneous twist deformation of the director field 
has been found in planarly aligned nematics \cite{Plaut}.
In homeotropic system with an applied magnetic field one has the
analogous situation, but here the homogeneous mode corresponds to a
rotation of the in-plane director without (or with small) twist
\cite{RHKP_PRL96}.

In comparison with the case of uniform director alignment (planar
or homeotropic)
only few works were devoted to EC in nematics with more complex
director distribution, e.g., in hybrid aligned nematics (HAN)
\cite{ADS_MCLC95,HKP_JPII95}.
Here below onset the director interpolates from the planar
alignment at one confining plate to homeotropic orientation
at the other one.
For this geometry one may also expect to find a transition to ARs
above threshold.
In this paper we present the results of an experimental and theoretical study of
the onset of electroconvection in hybrid aligned nematic in conductive regime
under a.c. voltage.
Below the Lifshitz frequency a secondary instability leading to drifting abnormal
rolls was found for the first time.

\vspace{24pt}
\noindent
{\bf EXPERIMENTAL}
\vspace{12pt}

\noindent
The nematic liquid crystal MBBA was sandwiched between two parallel
glass plates with transparent electrodes.
In order to achieve a uniform planar alignment, one  electrode was
rubbed in one direction.
Homeotropic alignment at the other electrode was obtained spontaneously
after cleaning of the electrode surface with alcohol.
The electrodes were separated by mylar spacers with thickness 
$d=40$~$\mu$m.
The lateral size of the cell was $10$ mm $\times$ $20$ mm, so that the 
aspect ratios of the cell were $250$ and $500$.
The temperature of the cell was kept at $T=25 \pm 0.1$ $^\circ$C.
The a.c. voltage applied across the NLC layer was 
generated by the digital wave synthesizer card WSB-100 (Quatech). 
The convection patterns were observed with polarising 
microscope and 
images were taken with a CCD-camera and digitized by frame-grabber 
DT3155 with resolution of $756 \times 581$ pixel and $256$ gray scale 
levels. 
The cut-off frequency of the NLC sample was $f_c \sim 60$ Hz.
The spatial periodicity of domain patterns was determined from the 
Fourier transform of images taken at threshold.

\vspace{24pt}
\noindent
{\bf RESULTS AND DISCUSSION}
\vspace{12pt}

\begin{figure}
\begin{center}
\vspace*{0.1cm}
\hspace*{-0.5cm}
\epsfxsize=11.0cm
\epsfbox{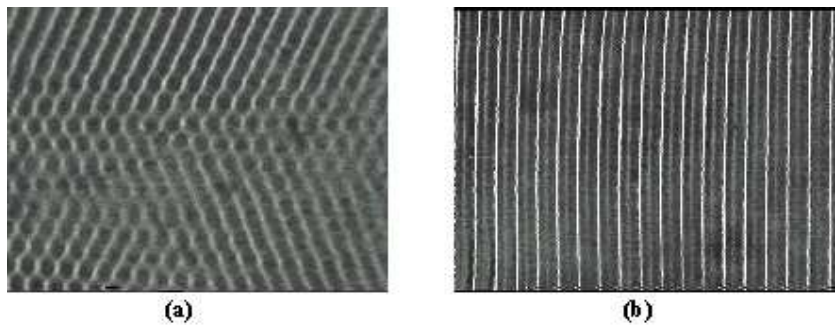}
\end{center}
\vspace*{-0.5cm}
\caption{Typical images of drifting rolls at onset in a HAN cell:
oblique rolls at $f=10$ Hz, $U=6.3$ V (a);
normal rolls at $f=30$ Hz, $U=8.5$ V (b).}
\label{fig1}
\end{figure}

\noindent
At the onset of EC ($U_c$) drifting rolls 
oriented obliquely or perpendicularly (normal) to the $x$ axis,
depending on a.c. frequency $f$ were 
observed (Fig.\ref{fig1}).

Below  the Lifshitz frequency $f_L$  oblique rolls arise 
at first at the edges of the NLC cell so that the zig-rolls appear near one 
edge and zag-rolls near another one. Superposition of zig- and zag-rolls
usually leads to the grid pattern formation in some places of NLC 
sample in the subcritical regime [Fig.\ref{fig1}(a)].
This pattern is unstable and above onset oblique rolls become preferable.

\begin{figure}
\begin{center}
\vspace*{0.1cm}
\hspace*{-1cm}
\epsfxsize=9.0cm
\epsfbox{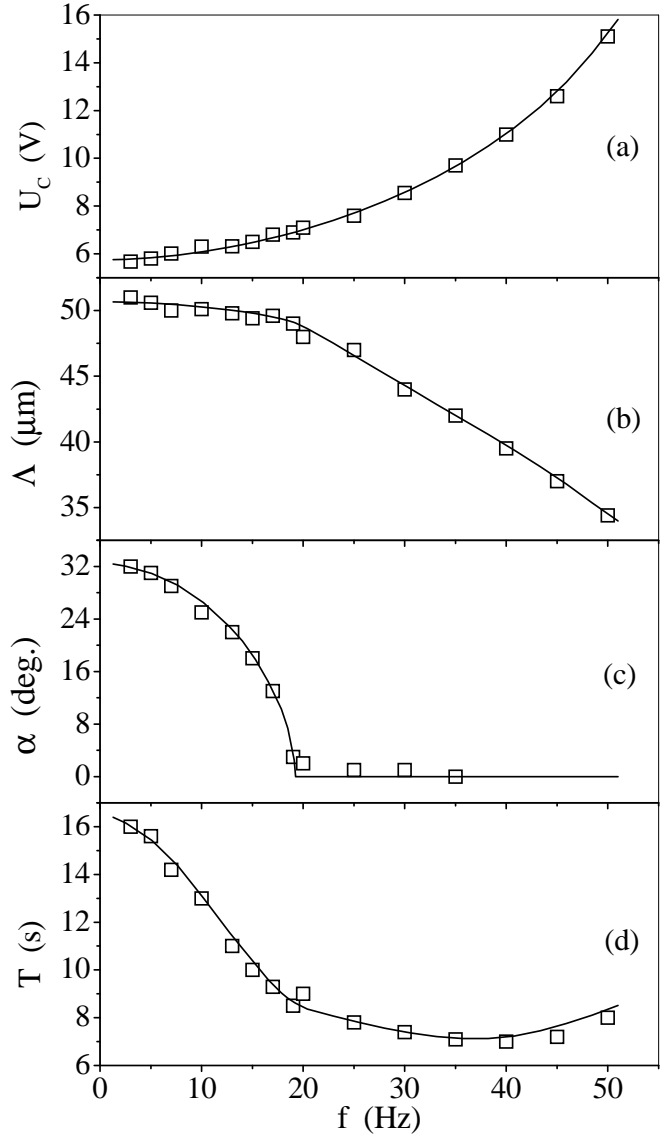}
\end{center}
\vspace*{-0.5cm}
\caption{Threshold voltage $U_c$ (a), roll period $\Lambda$ (b),
roll oblique angle $\alpha$ (c) and drift roll period $T$ versus
a.c. voltage frequency $f$.}
\label{fig2}
\end{figure}

Increasing the a.c. frequency a transition from oblique 
rolls to normal rolls was found at the Lifshitz frequency 
$f_L \simeq 19.8$~Hz.
The drifting rolls were observed in all 
range of a.c. frequency and the drift direction 
is determined by the sign of the director gradient
along the normal to the NLC layer in the initial hybrid state.

We have measured the threshold voltage $U_c$, roll period 
$\Lambda$, oblique angle $\alpha$ and drifting period $T$ as function
of the a.c. frequency $f$.
These experimental data are in a good agreement with the results of
linear stability analysis (Fig.\ref{fig2}).

In our calculations we used the standard set of nematodynamic equations 
\cite{BZK_JP88,HKP_JPII95}. 
Below onset of convection one has the director configuration distorted across
the layer due to the opposite boundary conditions and no material flow 
is excited. The linear stability analysis of this state is based on a Galerkin 
method (see \cite{HKP_JPII95} for details). 
Material parameters of MBBA at $25$ $^\circ$C \cite{HKP_JPII95} were 
used with the only exceptions the values of conductivities
$\sigma_{||}/\sigma_{\perp}=1.71$ and
$\sigma_{\perp}=0.55 \cdot 10^{-8}$ ($\Omega \cdot$m)$^{-1}$
which were fitted from the experimental data on the threshold voltage $U_c$ 
at low frequency of a.c. voltage and the cut-off frequency $f_c$, 
respectively.

In order to reveal the direction of the mean drift velocity, small tracer particles
($2-4$ $\mu$m in diameter) were immersed in the nematic.
The observations show that the particles move on closed trajectories within EC rolls
in the plane perpendicular to the roll axis and also drift together with rolls.
We found that the direction of the particle motion is not parallel to the wavevector
for the oblique roll regime.
In particular, for the applied voltage $U=6.3$ V at $f=10$~Hz
the direction of the mean drift velocity has an angle $\sim 20^\circ$
with respect to the wavevector.
This is also in agreement with the theoretical calculations. 

\begin{figure}
\begin{center}
\vspace*{0.3cm}
\hspace*{-1cm}
\epsfxsize=7.5cm
\epsfbox{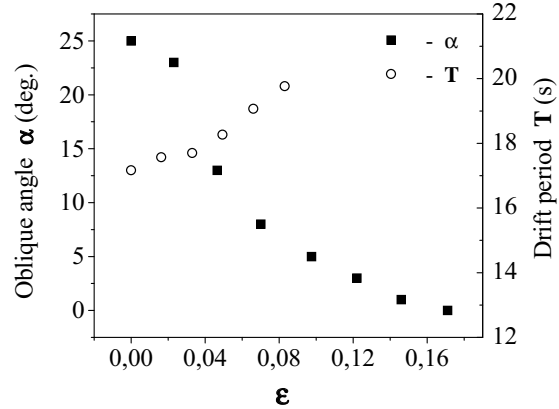}
\end{center}
\vspace*{-0.5cm}
\caption{The dependence of the roll angle $\alpha$ and
drift period $T$ on the control parameter $\epsilon$ for a.c.
frequency $f=10$ Hz.}
\label{fig3}
\end{figure}

With increasing voltage above the threshold $U_c$ the angle of
obliqueness decreases continuously and 
vanishes at some critical value of the control parameter 
$\epsilon_{AR} = (U_{AR}^2-U_c^2)/U_c^2$.
The dependence of the roll angle $\alpha$ on the control 
parameter $\epsilon$ for a.c. frequency $f=10$ Hz is presented
in Fig.\ref{fig3}.
The drift period $T$ of oblique rolls increases slightly with applied
voltage (Fig.\ref{fig3}).
Further increasing of the applied voltage leads to the appearance of
defects and weak turbulence is developed.

The polarised-optical analysis shows that at $\epsilon_{AR}$ the director is 
twisted away from the plane of initial orientation ($x-z$ plane).
In the case of hybrid cell, when the 
incident light enters into the lower 
electrode with planar director alignment,  polarisation of the 
transmitted light is parallel to the director projection onto the
$x-y$ plane (${\bf \hat c}$ director) in the bulk of the NLC.
Therefore it is quite easy to determine the effective twist angle of the
director $\phi$ by rotation of the analyser by an angle that corresponds
to maximum contrast of the intensity profiles along the line 
perpendicular to the rolls.
We found that the twist angle $\phi$ practically does not change
beyond $\epsilon_{AR}$.
For example, at $f=10$ Hz
and $\epsilon>\epsilon_{AR}=0.17$ one has
$\phi \simeq \pm 45^\circ$. 

\begin{figure}
\begin{center}
\vspace*{-0.1cm}
\hspace*{-0.5cm}
\epsfxsize=11.0cm
\epsfbox{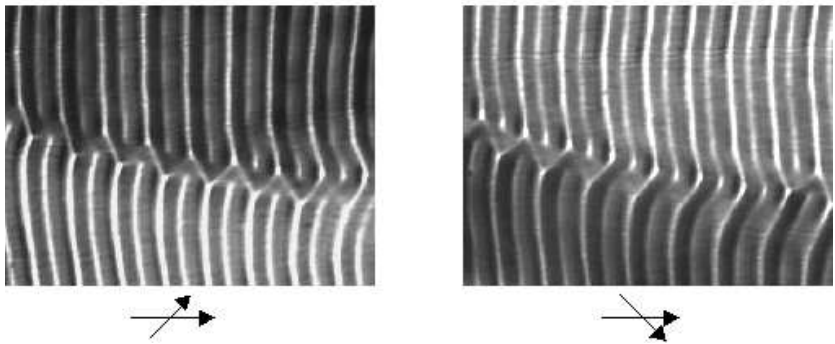}
\end{center}
\vspace*{-0.8cm}
\caption{Snapshots of abnormal rolls taken for two positions of
analyser with polariser along $x$ axis (arrows) demonstrated two
symmetry-degenerated ${\bf \hat c}$ director orientations ($\pm \phi$).}
\label{fig4}
\end{figure}

Thus, there is a symmetry 
breaking of the in-plane director, i.e., 
one has a kind of transition to {\em drifting abnormal rolls}.
The system
is divided into areas consisting of drifting rolls with two 
symmetry-degenerate ${\bf \hat c}$-director orientations ($\pm \phi$), 
which are separated from each other by moving domain walls (Fig.\ref{fig4}).

\vspace{24pt}
\noindent
{\bf CONCLUSION}
\vspace{12pt}

\noindent
We have shown that the results of linear stability analysis and the 
experimental data on the threshold characteristics of EC in the
hybrid aligned nematic under a.c. voltage are in a good quantitative agreement.

In contrast to the case of planarly aligning nematics,
the primary bifurcation is non-stationary and drifting 
rolls are observed at the onset.
The drift direction is determined by the sign of
the director gradient along the normal to the NLC layer in the
initial hybrid state.
A particularly interesting feature is that the direction of the
mean drift velocity below the Lifshitz frequency does not coincide
with wavevector.

The transition from drifting oblique rolls to drifting abnormal rolls
was found below Lifshitz frequency with increasing applied voltage.
The study of the dynamics of domain walls separating abnormal rolls with
two degenerate ${\bf \hat c}$-director orientations is in progress.

\vspace{12pt}
\noindent
{\bf Acknowledgments}

\noindent
We thank L. Kramer for fruitful discussions and critical reading
of manuscript.
We wish to acknowledge the hospitality of the University of Bayreuth. 
Financial support from DFG Grants Kr-690/14-1, 436-RUS-113/220
and INTAS Grant 96-498 is gratefully acknowledged. V.D. is also
grateful to the INTAS for fellowship grant YSF 99-4036.

\setlength{\baselineskip}{12pt}

\end{document}